\documentclass[american,aps,pra,superscriptaddress,reprint,showpacs]{revtex4-1}
\usepackage[T1]{fontenc}
\usepackage[latin9]{inputenc}
\usepackage{babel}
\usepackage{amsthm}
\usepackage{amsmath}
\usepackage{bm}
\usepackage{tabularx}
\usepackage{amssymb}
\usepackage{graphicx}
\usepackage{tikz}
\usepackage{float}
\usepackage{subfigure}
\usepackage{mathrsfs}
\usepackage{epstopdf}
\usepackage{fancyhdr}
\usepackage{multirow}
\usepackage[normalem]{ulem} 

\usepackage[unicode=true,pdfusetitle, bookmarks=true,bookmarksnumbered=false,bookmarksopen=false,
 breaklinks=false,pdfborder={0 0 0},backref=false,colorlinks=false]
 {hyperref}
\hypersetup{colorlinks,linkcolor=myurlcolor,citecolor=myurlcolor,urlcolor=myurlcolor}

\makeatletter
\@ifundefined{textcolor}{}
{%
 \definecolor{BLACK}{gray}{0}
 \definecolor{WHITE}{gray}{1}
 \definecolor{RED}{rgb}{1,0,0}
 \definecolor{GREEN}{rgb}{0,1,0}
 \definecolor{BLUE}{rgb}{0,0,1}
 \definecolor{CYAN}{cmyk}{1,0,0,0}
 \definecolor{MAGENTA}{cmyk}{0,1,0,0}
 \definecolor{YELLOW}{cmyk}{0,0,1,0}
 }
\theoremstyle{plain}

\ifx\proof\undefined
\newenvironment{proof}[1][\protect\proofname]{\par
\normalfont\topsep6\p@\@plus6\p@\relax
\trivlist
\itemindent\parindent
\item[\hskip\labelsep
\scshape
#1]\ignorespaces
}{%
\endtrivlist\@endpefalse
}
\providecommand{\proofname}{Proof}
\fi

\usepackage{times}
\usepackage{txfonts}
\usepackage{braket}
\usepackage{colortbl}
\definecolor{myurlcolor}{rgb}{0,0,0.7}

\makeatother

\providecommand{\theoremname}{Theorem}

\usepackage{times}

\newcommand{\bq}{\begin{eqnarray}}
\newcommand{\eq}{\end{eqnarray}}
\newcommand{\be}{\begin{equation}}
\newcommand{\ee}{\end{equation}}
\newcommand{\ei}{\mathrm{e}}
\newcommand{\tr}{\operatorname{tr}}

\newcommand{\abs}[1]{\left| #1 \right|}

\newcommand{\nkb}[2]{\xout{#1}{\ \color{blue}#2}} 
\newcommand{\ankb}[2]{#2} 

\newcommand{\alams}[2]{#2}

\begin{document}

\title{Maximizing complementary quantities by projective measurements}

\begin{abstract}
In this work we study the so-called quantitative complementarity quantities. We focus in the following physical situation: two qubits ($q_A$ and $q_B$) are initially in a maximally entangled state. One of them ($q_B$) interacts with a $N$-qubit system ($R$). After the interaction, projective measurements are performed in each of the qubits of $R$, in a basis that is chosen after independent optimization procedures: maximization of the visibility, the concurrence and the predictability. For a specific maximization procedure, we study in details how each of the complementary quantities behave, conditioned on the intensity of the coupling between $q_B$ and the $N$ qubits. We show that, if the coupling is sufficiently ``strong'', independent of the maximization procedure, the concurrence tends to decay quickly. Interestingly enough, the behavior of the concurrence in this model is similar to the entanglement dynamics of a two qubit system subjected to a thermal reservoir, despite that we consider finite $N$. \textcolor[rgb]{0.00,0.00,0.00}{However the visibility shows a different behavior: its maximization is more efficient for stronger coupling constants.} Moreover, we investigate how the distinguishability, or the information stored in different parts of the system, is distributed for different couplings.
\end{abstract}

\author{Leonardo A. M. Souza}
\email{leonardoamsouza@ufv.br}
\affiliation{Universidade Federal de Vi{\c c}osa - Campus Florestal,
LMG818 Km6, Minas Gerais, Florestal 35690-000, Brazil}

\author{Nadja K. Bernardes}
\affiliation{Departamento de F\'{i}sica, Universidade Federal de Minas Gerais,
Belo Horizonte, Caixa Postal 702, 30161-970, Brazil}

\author{Romeu Rossi Jr.}

\affiliation{Universidade Federal de Vi{\c c}osa - Campus Florestal,
LMG818 Km6, Minas Gerais, Florestal 35690-000, Brazil}

\date{\today}
\maketitle

\section{Introduction}

The wave-particle duality is an alternative statement of the complementarity principle, and it establishes the relation between \ankb{corpuscular and undulatory}{the corpuscular and the ondulatory} nature of quantum entities \cite{Bohr1928}. It can be illustrated in a two-way interferometer, where the apparatus can be set to observe \ankb{particle}{the particle} behavior when a single path is \ankb{taken}{taken,} or wave-like behavior, when the impossibility to define a path is shown by the interference. A modern approach to the wave-particle duality includes quantitative relations between quantities that represent the possible \textit{a priori} knowledge of the which-way information (\ankb{predictability}{predicability}) and the ``quality'' of the interference fringes (\ankb{Visibility}{visibility}). Several publications in the literature \cite{Bohr1928, Wootters1979, Summhammer1987, Greenberger1988, Mandel1991} contributed to the formulation of the quantitative analysis of the wave-particle duality. For a bipartite system\ankb{ entanglement, the quantum correlations between each part, can play a role. Such correlations can}{, entanglement can} give an extra which-way (path) information about the interferometric possibilities. The quantitative relations for systems composed by two particles were \ankb{}{extensively} studied in \cite{Jaeger1993, Jaeger1995, Englert1996, Englert2000, Scully1989, Scully1991, Mandel1995, Tessier2005, Jakob2010, Miatto2015,Bagan2016, Coles2016}. Therefore, \ankb{understand}{understanding} the behavior of such quantities, in various regimes and situations, is essential to answer fundamental and/or technological questions of the quantum theory \cite{Greenberger1999}.

\alams{The Complementarity quantities can present interesting dynamical behaviors,}{Concerning the study of the dynamical behavior of complementarity quantities,} an example is the so-called \textit{quantum eraser}, \alams{where an increase or preservation of the visibility of an interferometer experiment is caused when the ``which-way'' information is erased.}{\ankb{ i.e. an increasing or preservation of the \ankb{Visibility}{visibility} in an interferometric scheme (or the ``erasure'' of the which-way information probably stored in the initial state).}{ where an increase or preservation of the visibility of an interferometer experiment is caused when the which-path information is erased.}} Since its proposal \cite{Scully1982}\ankb{ this phenomena}{, it} has \ankb{investigated it carefully,}{been investigated carefully} both theoretically and experimentally (see for example Refs. \cite{Englert2000, Scully1991, Mandel1995, Storey1994, Wiseman1995, Mir2007, Luis1998, Busch2006, Rossi2013, Walborn2002, Mir2007, Teklemariam2001, Teklemariam2002, Kim2000, Salles2008, Heuer2015}). In \ankb{a recent work }{Ref.~}\cite{Rossi2013}, the authors explore the quantum eraser problem in \ankb{multipartite}{a multipartite} model\ankb{, where two cavities ($q_A+q_B$), which will be taken as a two qubit system $A + B$, in an initial maximally entangled state (and therefore with zero \ankb{Visibility}{visibility}), couple through a Jaynes-Cummings Hamiltonian to $N$ two-level atoms (we will call the global system as $q_A + q_B + R$, where all the individual systems are qubits)}{. Initially a bipartite qubit system is prepared in a maximally entangled state and interacts with $N$ other qubits. This model can be implemented considering the qubits of interest the cavity modes of two cavities and the $N$ qubits as two-level atoms}. In this work \cite{Rossi2013}, an increase of visibility is achieved by performing appropriate projective measurements. An intrinsic relation between the complementarity quantities and the performed measurements is outlined: since \ankb{they}{the measurements} were made in order to obtain an \ankb{increasing}{increase} of the \ankb{Visibility}{visibility}, the remaining quantities (Entanglement as measured by the concurrence, and the predictability) must obey a ``complementary'' behavior. In that case, \ankb{Visibility}{visibility} and predictability increases, and entanglement decreases, since the measurements are made in order to \ankb{establishes}{establish} the quantum eraser. In Reference \cite{Rossi2013} only the maximization of the visibility was considered, in the present work we extend the analysis and consider maximization of predictability, visibility and concurrence. Also, in the previous work \cite{Rossi2013}, only one value of the coupling constant was considered. In this contribution we consider a second coupling regime that allows for the comparison between stronger and weaker interactions.

Some questions may arise from the analysis presented in \cite{Rossi2013}: how is the behavior of the \ankb{Visibility}{visibility}, predictability and Entanglement, for different strengths of the coupling between the cavities and the $N$ atoms? \ankb{There are any difference in this behavior,}{Is there any difference in this behavior} if one measure the qubits in order to maximize another complementarity quantity? For finite $N$, could \alams{this behavior}{the behavior of entanglement} resemble the reservoir (dissipative) limit? Moreover, one can think about a three-part control scheme: initially parts $A$ and $B$ possesses a maximally entanglement state, constituted by two qubits $q_A$ and $q_B$, respectively. A third part $R$ may have, in principle, \emph{full} control of a group of $N$-qubits (each one we call as $q_i$); i.e. $R$ may control: (i) the initial state of each qubit $q_i$, (ii) the interaction strength between $q_i$ and $q_B$ and (iii) the measurement basis where each $q_i$ could be projected by $R$. Here we will focus in the control of item (iii), therefore the initial state of all $q_i$ and the coupling strenght will be fixed for each realization of the scheme. Thus, part $R$ is able to control which complementarity quantity of part $A$ they ($A$ and $R$) would like to maximize. For instance, if $A$ and $R$ desire that $q_A$ is in a superposition state, $R$ can choose which basis he/she will project each qubit in order to accomplish the task (quantum eraser task \cite{Rossi2013}). However, now $R$ and $A$ are able to choose another complementarity quantity: if they would like to obtain and/or maintain an Entangled state between $A$ and $B$, $R$ may project each $q_i$ in a basis chosen in order obtain an state nearly maximally Entangled (the same idea follows for the predictability). More than that, since part $R$ can adjust the strenght of the interaction between $q_B$ and $q_i$, he/she can study what is the best option of coupling to do each task (together with the freedom to choose the basis of projection). In that way, parts $A$, $B$ and $R$ are able to study in details the behavior of the complementarity quantities, for a variety of conditions.

In the present work we answer the questions and provide a useful tool to implement the control scheme mentioned above, considering a similar model\ankb{: the interaction between}{ compound by} two entangled qubits, $q_{A}$ and $q_{B}$, and a third system ($R$) which is composed by $N$ qubits. \ankb{They interact, one at the time, with the qubit B. The $N$ qubits of $R$ can be measured after the interaction.}{Each qubit of $R$ interacts one at a time with only qubit $B$ and can be projectively measured afterwards.} It is well known that \ankb{when the interaction time $t\rightarrow 0$ and on the coupling strength $g\rightarrow \infty$ the subsystem}{in the limit that the interaction time $t\rightarrow 0$ and that the coupling strength $g\rightarrow \infty$, the system} $R$ will play the role of a reservoir \cite{Carmichael1999, Breuer2007, Jacobs1998}. As it is possible to measure each qubit of system $R$ after the interaction, we can control the evolution of $q_{A}$ and $q_{B}$, induced by the interaction with $R$, by selecting an adequate sequence of results of measurements performed in the qubits of $R$. Such control would allow us to make $q_{A}$ and $q_{B}$ approach a chosen asymptotic state. This scheme can be implemented in cavity-QED system, where $q_{A}$ and $q_{B}$ would be cavity modes, prepared in an entangled state with one excitation, and $N$ two level atoms, interacting with the cavities one at the time, would play the role of the qubits that compose the system $R$. We consider the complementarity quantities \cite{Jakob2010} concurrence, predictability and \ankb{Visibility}{visibility} to guide the manipulation over $R$ and to quantify the information present in each subsystem. \ankb{To manipulate the evolution we consider three kinds of sequences of experimental results: The first maximizes the Visibility, the second maximizes the predictability and the third maximizes the concurrence.}{Each quantity is maximized by a different set of projective measurements on $R$.} We also consider two regimes for the strength of the coupling constant $g$ between $q_B$ and each qubit of $R$, $g= \frac{1}{4}$ and $g= 4$. \ankb{We perform a numerical calculation in order to find the right sequence of measurement that lead to the maximization of a given quantity.}{} \alams{We show that for $g= \frac{1}{4}$ it is possible to manipulate the evolution to make the subsystem of interest $q_{A}+q_{B}$ to approach a chosen asymptotic state. For $g T = 2 \pi \times 4$, the subsystem $q_{A}+q_{B}$ always tends to a state with no excitation in $q_{B}$, independently of the chosen sequence of measurement results in $R$. Therefore, when the coupling constant increases, the result of the interaction between the system of interest and $R$ is similar to the interaction with a thermal reservoir. We consider also the case with no maximization, assuming that \ankb{the}{} all measurements are made in the same basis while we observe the complementarity quantities behavior. Finally we show how the information is distributed over the global system, and how this distribution changes after the measurements in $R$.}{\textcolor[rgb]{0.00,0.00,0.00}{We show that for $g = \frac{1}{4}$ it is possible to maximize the concurrence of the subsystem $q_A+q_B$, while for $g T= 2 \pi \times 4$ (i.e. $g = 4$) the concurrence decays quickly and the maximization is not possible. When the coupling constant increases, the behavior of concurrence is similarly to the one expected if the system $R$ had the properties of a thermal reservoir. However, the visibility shows a different behavior, its maximization is more efficient for $g = 4$. This behavior is caused by the different which-way information distribution produced by the interactions with $g = 4$ and $g=\frac{1}{4}$, as we shown in the last section. Numerical calculation shows that, for $g = 4$, the first two qubits of $R$ retain a large amount of which-way information, that was initially present in $q_B$. When measurements that maximizes the 
visibility are performed, the which-way information is erased, and the visibility of $q_A$ increases 
quickly. For $g=\frac{1}{4}$ the which-way information is distributed almost equally among the qubits of $R$, therefore less information is erased and consequently measurements that maximize the visibility are less efficient.}}

The paper is organized as \ankb{follow}{follows}: in section \ref{model} we \ankb{present the model in details, including the complete dynamics of the global system and}{briefly review the model and the definition of} the principal quantities studied: \ankb{Visibility}{visibility}, concurrence and predictability. \ankb{We also study in this section}{Moreover, we analyze} the distinguishability between different parts of the global system\ankb{ ($q_A$ and $q_B$, $q_A$ and the i-esim qubit of $R$)}{}. In subsection \ref{digression} we briefly review the case where $q_A+q_B$ interacts in a dissipative reservoir, and how the complementarity quantities behave in this case. Section \ref{results} shows how we implement the projective measurements in $R$, and present our results and discussions for the behavior of the complementarity quantities and for the variation of distinguishability (after and before the measurements). In section \ref{conclusion} we conclude our work.

\begin{figure}[h]
\centering
{\includegraphics[scale=0.32]{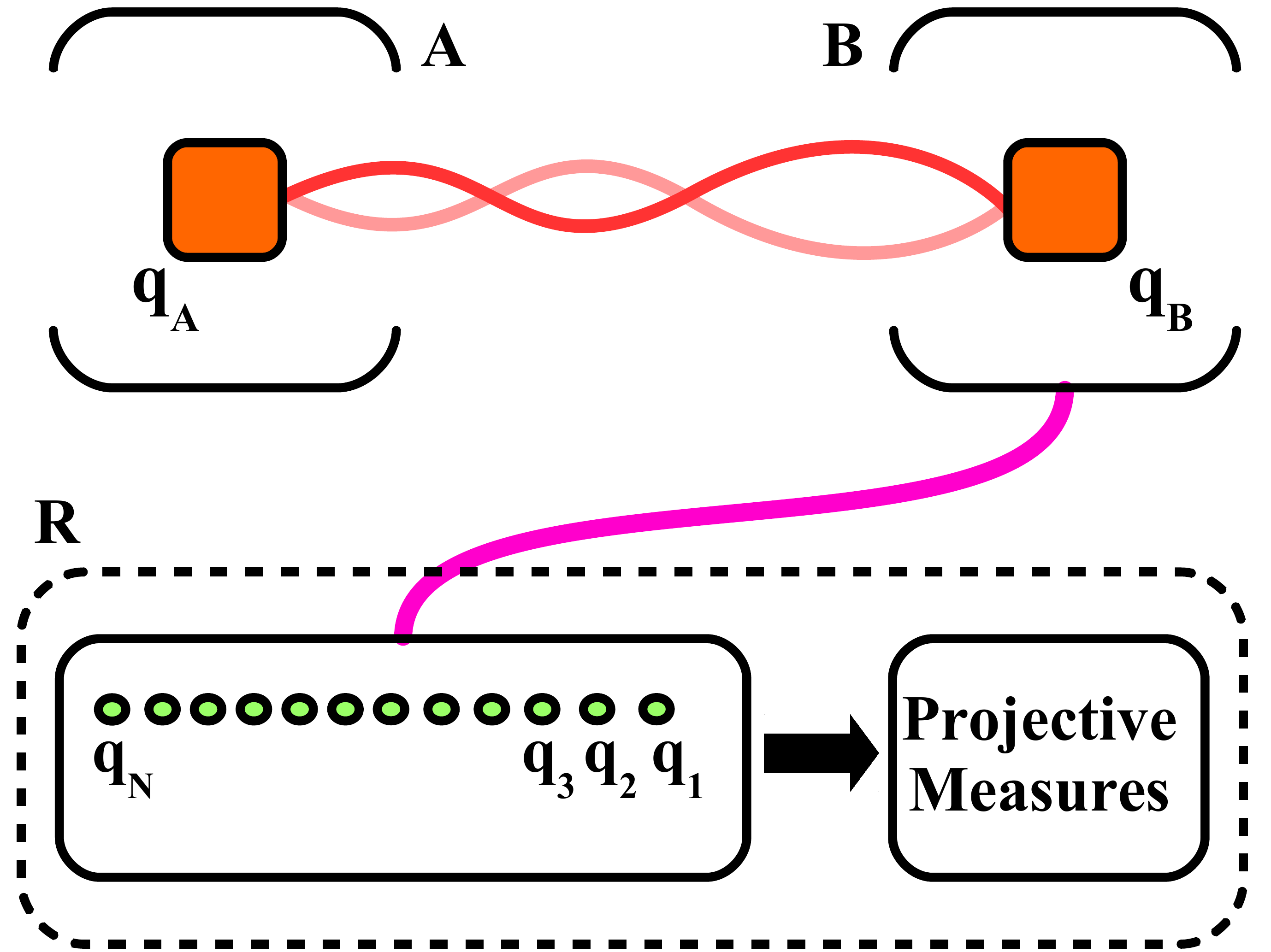}}
\caption{(Color online) A schematic figure of our proposal. The qubits $q_A$ and $q_B$ are initially Entangled (orange squares), and part $R$ (which contains $N$ qubits, represented by green circles) interacts with $q_B$, throught the Hamiltonian indicated in the text (represented by the pink curve). The projective measures follows after $R$ decides which complementarity quantity of $A$ and $B$ will be maximized.}
\label{scheme}
\end{figure}

\section{Model and definitions}\label{model}

Let us consider that initially qubits $q_A$ and $q_B$ were prepared in the entangled state $\ket{\psi(0)} = \frac{1}{\sqrt 2} (\ket{0_A 1_B} + \ket{1_A 0_B})$ and a third system $R$ composed by \ankb{$N$-qubits}{$N$ qubits}, prepared all in the ground state $\ket 0$. Each qubit of $R$ interacts one at a time with $q_B$ and after a sequence of these interactions, the information, initially stored in $q_A$ and $q_B$, will be distributed over the \ankb{$N$-qubits of $R$}{$N$ qubits of $R$} and qubits $q_A$ and $q_B$. As an example of our interaction model, consider the following dynamics governing the interaction of an atom (between a total of $N$ atoms) and a cavity ($q_B$) (see Figure \ref{scheme}). The Hamiltonian that gives the interaction between the $k$-th atom and $q_B$ is $\hat{H}^{(k)}= \omega \hat{b}^{\dagger }\hat{b}+ \frac{\omega}{2}\hat{\sigma}_{z}^{(k)}+ g (\hat{b}^{\dagger}\hat{\sigma}_{-}^{(k)}+\hat{b}\hat{\sigma}_{+}^{(k)}),$ where $\hat{b}^{\dagger }$ ($\hat{b}$) corresponds to  the creation (annihilation) operator for $q_B$, $\omega $ their transition frequency, $\hat{\sigma}_{z}^{(k)}=|1^{(k)}\rangle\langle 1^{(k)}|-|0^{(k)}\rangle\langle 0^{(k)}|$, $\hat{\sigma}_{-}^{(k)}=|0^{(k)}\rangle\langle 1^{(k)}|$, $\hat{\sigma}_{+}^{(k)}=|1^{(k)}\rangle\langle 0^{(k)}|$, \ankb{}{and} $g$ the coupling constant for the interaction between the $k$-th qubit of $R$ and $q_B$. As the initial state has one excitation and the Hamiltonian preserves the excitation number, the states of mode $B$ can be written in the basis $\lbrace\vert 0 \rangle , \vert 1 \rangle\rbrace$. Although constant in each preparation, we let the parameter $g$ free in order to quantify the strength of the interaction, since we will analyze different coupling regimes and the correspondent behavior of the Complementarity quantities. In this model, $\left|0^{(k)} \right\rangle$ and $\left|1^{(k)}\right\rangle$ stand for the levels $0$ and $1$ of the $k$-th interacting atom, respectively.

After $n$ qubits of $R$ have interacted with $q_B$ (note the difference between $N$, the total number of qubits that are able to interact, and $n$, the number of qubits that will interact at a given time) the global system is left in the state \cite{Rossi2013}
\bq \ket{\psi^{(n)}} &=&
\frac{1}{\sqrt 2} \Big( a^n \ket{0_{A}} \ket{0_{res}} \ket{1_{B}} + \Gamma \ket{0_{A}}
\ket{1_{res}} \ket{0_{B}} + \nonumber \\ && + \ket{1_{A}} \ket{0_{res}} \ket{0_{B}} \Big), \label{psin}\eq
where $a = \cos\left( \frac{g T}{N} \right)$ and $b = - i \sin \left( \frac{g T}{N} \right)$, assuming the same interaction time between each qubit of $R$ and $q_{B}$, given by $\Delta
t=\frac{T}{N}$. To simplify the notation we define a normalized state with one excitation in subsystem $R$:  $\ket{1_{res}}=
\frac{1}{\Gamma} \left( a^{n-1} b \ket{n} + \ldots + a b \ket{2} + b\ket{1} \right)$\ankb{ with}{, where} $\Gamma^{2} = 1 - a^{2n}$\ankb{ and}{,} $\ket{i}= \ket{0_1 0_2 \ldots 0_{i-1} ~1_i ~ 0_{i+1} \ldots 0_n}$ represents a state with an excitation in the $i$-th qubit and $\ket{0_{res}} = \ket{0_1 0_2 \ldots 0_n}$. 

For a general pure two qubit state \cite{Englert1996, Englert2000, Jakob2010}:
$\ket{\Psi} = \gamma_1 \ket{00} + \gamma_2 \ket{01} + \gamma_3 \ket{10} + \gamma_4 \ket{11},$
the complementary quantities are defined in the following way. The concurrence, which is related to the quantum correlation between the parts, is given by $C(\ket{\Psi}) = 2 \abs{\gamma_1 \gamma_4 - \gamma_2 \gamma_3}$. The coherence between two orthogonal states gives the visibility, defined by $V = 2 \abs{\bra{\Psi}\sigma_{+}\ket{\Psi}} = 2 \abs{\gamma_1 \gamma_3^* + \gamma_2 \gamma_4^*}$. Besides, the predictability measures the knowledge if one of the parts is in state $\ket{0}$ or $\ket{1}$, $P= \abs{\bra{\Psi}\sigma_{z}\ket{\Psi}} = \abs{\abs{\gamma_3}^2+\abs{\gamma_4}^2- \left( \abs{\gamma_1}^2 + \abs{\gamma_2}^2 \right)}$. The distinguishability, or the ``measure of the possible which-path information that one can obtain'' in an interferometer setup \cite{Englert2000, Jakob2010}, is given by $D = \sqrt{C^2 + P^2}$. \alams{}{For completeness, we present here explicitly the global system state operator: \begin{widetext}
\bq \rho^{(n)} &=& \ket{\psi^{n}}\bra{\psi^{n}} = \frac{1}{2}
\Big( a^{2n} \ket{0_{A} ~ 0_{res} ~ 1_{B}} \bra{0_{A} ~ 0_{res} ~ 1_{B}} + \Gamma^2
\ket{0_{A} ~ 1_{res} ~ 0_{B}} \bra{0_{A} ~ 1_{res} ~ 0_{B}} + \ket{1_{A} ~ 0_{res} ~ 0_{B}} \bra{1_{A} ~ 0_{res} ~ 0_{B}} + \nonumber \\ && + a^n \Gamma^* \ket{0_{A} ~
0_{res} ~ 1_{B}} \bra{0_{A} ~ 1_{res} ~ 0_{B}} + a^n \ket{0_{A} ~ 0_{res} ~ 1_{B}}
\bra{1_{A} ~ 0_{res} ~ 0_{B}} + \Gamma \ket{0_{A} ~ 1_{res} ~ 0_{B}} \bra{1_{A} ~
0_{res} ~ 0_{B}} + h.c. \Big), \label{rhon} \eq where h.c. stands for the hermitian conjugate of the previous quantity. The distinguishability between $q_A$ and $q_B$ is obtained from the reduced state operator of subsystem $q_{A}+q_{B}$:
\bq
\rho_{q_{A}, q_{B}}^{(n)} =\tr_{res} \rho^{(n)} = \frac{1}{2} \Big[ \ket{0_{A}} \bra{0_{A}} \Big( a^{2n}
\ket{1_{B}} \bra{1_{B}} + \Gamma^2 \ket{0_{B}} \bra{0_{B}} \Big) + \ket{1_{A}} \bra{1_{A}} \ket{0_{B}}
\bra{0_{B}} + a^n \ket{0_{A} 1_B} \bra{{1_{A}} 0_B} + h.c. \Big], \eq \end{widetext} as: \bq D_{q_{A},q_{B}}^{(n)} &=& \tr \Big\{ \abs{\frac{1}{2} \Big( a^{2n}
\ket{1_{B}} \bra{1_{B}} + \Gamma^2 \ket{0_{B}} \bra{0_{B}} \Big) - \frac{1}{2} \ket{0_{B}}
\bra{0_{B}}} \Big\} \nonumber \\ &=&  \frac{1}{2} (a^{2n}
+ \abs{\Gamma^2 - 1}) = a^{2n}. \label{DnqAqB}\eq}

\subsection{Continuous Limit - A digression}\label{digression}

Defining $k = g^2 \frac{T}{N},$ one can write \cite{Carmichael1999, Breuer2007, Jacobs1998} $a = \cos \left(\sqrt{\frac{k T}{N}} \right),$ where $T$ is the total time of interaction between $q_B$ and $R$ \ankb{(composed by $N$-qubits)}{}. We consider that the interaction time between each qubit of $R$ and $q_B$ is equal, given by $\Delta
t=\frac{T}{N}$. We also assume that $0<\Delta t<\pi/2g.$ After $N$ interactions the reduced state in the subsystem $q_A$ is a statistical mixture $\rho_{A}=\frac{1}{2}\left(\vert 0_{A} \rangle \langle 0_{A} \vert + \vert 1_{A} \rangle \langle 1_{A} \vert\right)$, therefore $V_{q_A} = 0$ and $P_{q_A} = 0$. The concurrence can be calculated from the reduced state of the subsystem $q_A, q_B$ and is given by $C_{q_A, q_B} = a^N$. The limit $N \rightarrow \infty$ (and consequently $g \rightarrow \sqrt{\frac{N}{T}} \rightarrow \infty$; $\Delta t \rightarrow dt$) is well known in the Literature \cite{Carmichael1999, Breuer2007, Jacobs1998} and it gives the reservoir limit (at a given temperature implicitly defined in $k$) of a qubit interacting with a Markovian pure dissipative reservoir. The term $a^n$ in \nkb{}{Eq.}\eqref{psin} is\ankb{,}{} in this limit\nkb{:}{} $\lim_{N \rightarrow \infty} a^N = \lim_{N \rightarrow \infty} \cos^N \sqrt{\frac{k T}{N}} = \ei^{-kT/2}$, and consequently the concurrence $C_{q_A, q_B} = \ei^{\frac{-k T}{2}}$ decays exponentially with $T$. 




\section{Results}\label{results}

\subsection{Complementarity quantities versus coupling intensity}

Similar to what was done in Ref.~\cite{Rossi2013}, let us now consider that, after $n$ interactions, the $i$-th qubit of $R$ is projected in the state: $\ket{M_i} = \alpha_i \ket{0_i} + \beta_i
\ket{1_i},$
where $\alpha_i = \cos \theta_i$ e $\beta_i = \ei^{i
\phi_i} \sin \theta_i$ (this measure can be done experimentally, see for example \cite{Haroche2006, Salles2008, Aguilar2014}). The vector state $\ket{M_i}$ is an eigenstate of the operator \ankb{$\hat{\sigma}_{i}=\vec{n}\cdot \vec{\sigma}_{i}$}{$\hat{\sigma}_{i}=\vec{n}\cdot \vec{\sigma}$} with $\vec{n}=\left(\sin2\theta_{i} ~ \cos2\phi_{i},~ \sin2\theta_{i}~\sin2\phi_{i},~\cos2\theta_{i}\right)$ and
\ankb{$\vec{\sigma}=\left(\sigma_{x,i},~\sigma_{y,i},~\sigma_{z,i}\right)$}{$\vec{\sigma}=\left(\sigma_{x},~\sigma_{y},~\sigma_{z}\right)$}, the Pauli matrices. One can, in principle, choose in which base ($\theta_i$ and $\phi_i$) the global state will be measured. \ankb{Considering}{Let us consider} projective measurements performed on the state \eqref{psin}, the projector is given by
$ \Pi = \Pi_1 \otimes \ldots
\otimes \Pi_n, $ with $\Pi_i = \mathbb{I}_1 \otimes \ldots \otimes
\mathbb{I}_{i-1} \otimes \ket{M_i}\bra{M_i} \otimes \ldots
\mathbb{I}_n.$ Notice that the projective measure $\Pi$ acts only on the subsystem $R$. 

After $n$ projective measurements the normalized global state vector is given by:
\bq \ket \psi^{(n,M)} &=& \frac{1}{N}\Big( \gamma_1 \ket{0_A} \ket M \ket{0_B} + \gamma_2 \ket{0_A} \ket{M} \ket{1_B} \nonumber \\ &&+ \gamma_3 \ket{1_A} \ket M \ket{0_B} \Big), \label{rhoreducedM}\eq where $\ket M = \ket{M_1} \ldots \ket{M_n}$, $N = \sqrt{\abs{\gamma_1}^2 +
\abs{\gamma_2}^2 + \abs{\gamma_3}^2},$ and \bq \gamma_1 &=&
\frac{1}{\sqrt 2} \left(b \sum_{i = 1}^{n} \left[ a^{i-1}
\frac{\beta_i}{\alpha_i}
\left( \prod_{j=1}^{n} \alpha_j \right) \right] \right),\nonumber \\ \gamma_2 &=& \frac{1}{\sqrt 2}
\left(a^n \prod_{i=1}^{n} \alpha_i \right), \nonumber \\
\gamma_3 &=& \frac{1}{\sqrt 2} \left( \prod_{i=1}^{n} \alpha_i \right).\eq 
The information carried by the qubits of $R$ are now embodied in the measurement outcomes $\theta_i$ and $\phi_i$. The complementarity quantities after the measurements are given by:
\bq V^{(n,M)}_{q_{A}} &=& \frac{2 \abs{\gamma_1 \gamma_3^*}}{N^2} \nonumber
\\ P^{(n,M)}_{q_{A}} &=& \frac{\abs{\abs{\gamma_3}^2
- \abs{\gamma_1}^2-\abs{\gamma_2}^2}}{N^2} \nonumber \\ C^{(n,M)}_{q_{A},q_{B}} &=& \frac{2 \abs{\gamma_2 \gamma_3}}{N^2}. \nonumber \label{complementarity} \eq Since the reduced state is pure \eqref{rhoreducedM}, the closure relation for complementarities \ankb{hold}{holds}: \be
\left(C^{(n,M)}_{q_{A},q_{B}}\right)^2 +\left(P^{(n,M)}_{q_{A}}\right)^2 +\left(V^{(n,M)}_{q_{A}}\right)^2 = 1. \label{complem1}\ee \ankb{Those}{These} quantities depend explicit on the coefficients $\theta_i$ and $\phi_i$ of $\ket M_i$. \alams{If one \ankb{perform}{performs} a measurement $P_i$ on the \ankb{i-th}{$i$-th} qubit for instance, he/she can in principle choose $\theta_i$ and $\phi_i$ \alams{arbitrarily}{so that the outcome of $P_i$ return the required information about}.}{In principle $\theta_i$ and $\phi_i$ can be chosen such that by performing a measurement $\Pi_i$, the complementary quantities will change accordingly.}

Concerning the Complementarity quantities, one can project the global state so that $V^{(n,M)}_{q_{A}}, P^{(n,M)}_{q_{A}}$ or $C^{(n,M)}_{q_{A}, q_B}$ acquire the maximum allowed values, after $n$ measurements on the qubits of $R$ have interacted with $q_B$. In \ankb{Reference \cite{Rossi2013}}{Ref.~\cite{Rossi2013}}, the authors studied a similar maximization procedure, \ankb{}{although} only for the \ankb{Visibility}{visibility} $V^{(n,M)}_{q_{A}}$\ankb{, in order}{. In order} to produce a multipartite quantum eraser\ankb{:}{,} the coefficients $\alpha_i$ and $\beta_i$ were chosen to obtain an \ankb{increasing}{increase} in the \ankb{Visibility}{visibility}, maintaining a standard value for the coupling parameter ($g T = 2 \pi$). Here we are interested in how each of the Complementarity quantities behaves, \ankb{if one change the coupling intensity $g$}{for different coupling intensities $g$'s} between $q_B$ and the $i$-th \ankb{qubits}{qubit} of $R$, while making projective measurements in each qubit of $R$. \alams{The measurement outcomes, i.e. the values of $\theta_i$ and $\phi_i$, are given by the following numerical maximization:}{The values of $\theta_i$ and $\phi_i$ were chosen by the following numerical simulation:} if the function to be maximized is the concurrence $C^{(n,M)}_{q_{A}, q_B}$, for example, the procedure gives the values of $\theta_i$ and $\phi_i$ that provides the maximum value of $C^{(n,M)}_{q_{A}, q_B}$, after $n$ qubits have interacted with $q_B$; then, in the possession of $\theta_i$ and $\phi_i$, one can evaluate $V^{(n,M)}_{q_{A}}$ and $P^{(n,M)}_{q_{A}}$. The same procedure is carried out in order to maximize the \ankb{Visibility}{visibility} or the predictability. Therefore, we have all the Complementarity quantities for each function to be maximized: $V^{(n,M)}_{q_{A}}, P^{(n,M)}_{q_{A}}$ or $C^{(n,M)}_{q_{A}, q_B}$.

\begin{figure}[h]
\centering
{\includegraphics[scale=0.43]{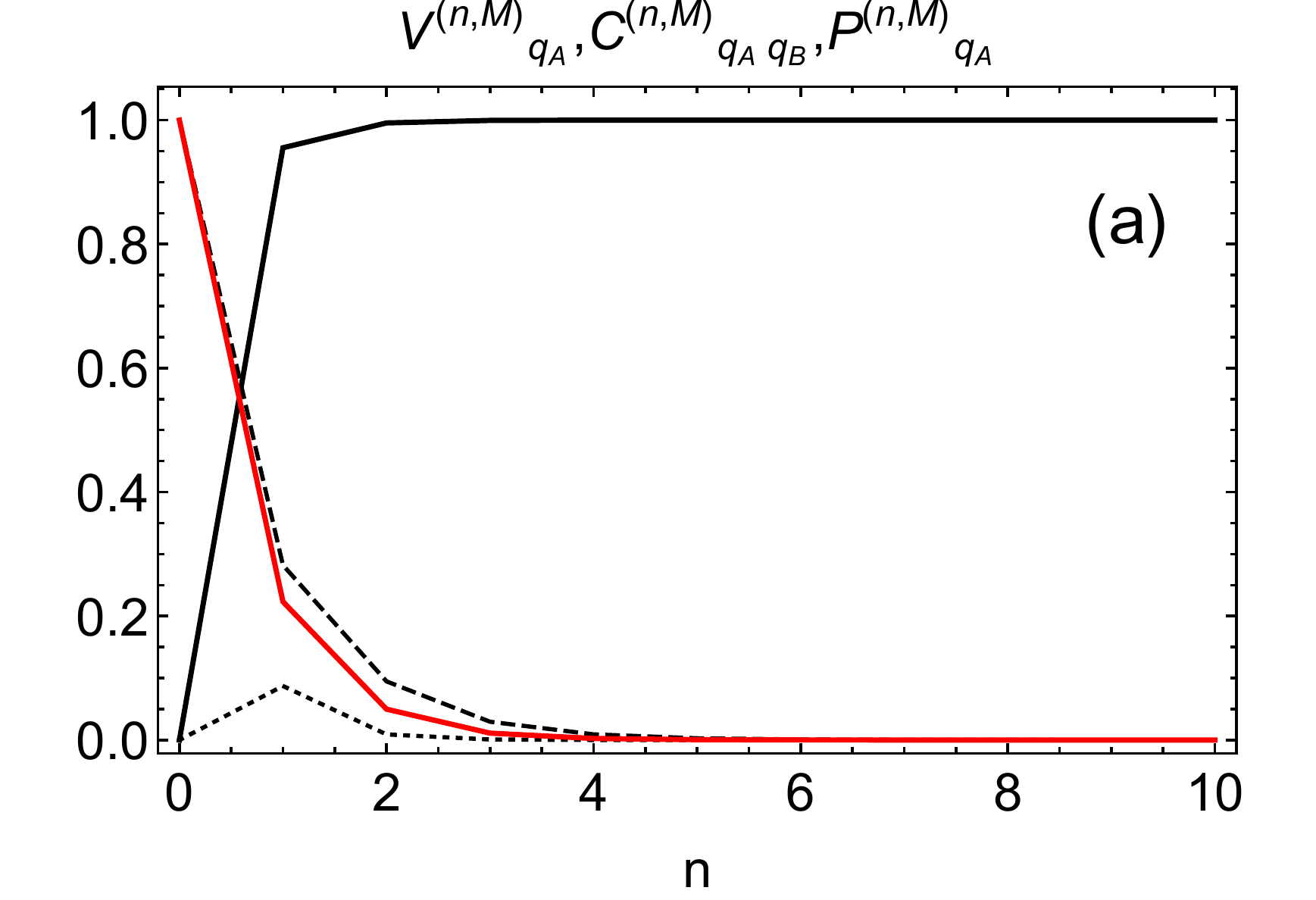} \label{maxvisa}}\hspace{0.5cm}
{\includegraphics[scale=0.43]{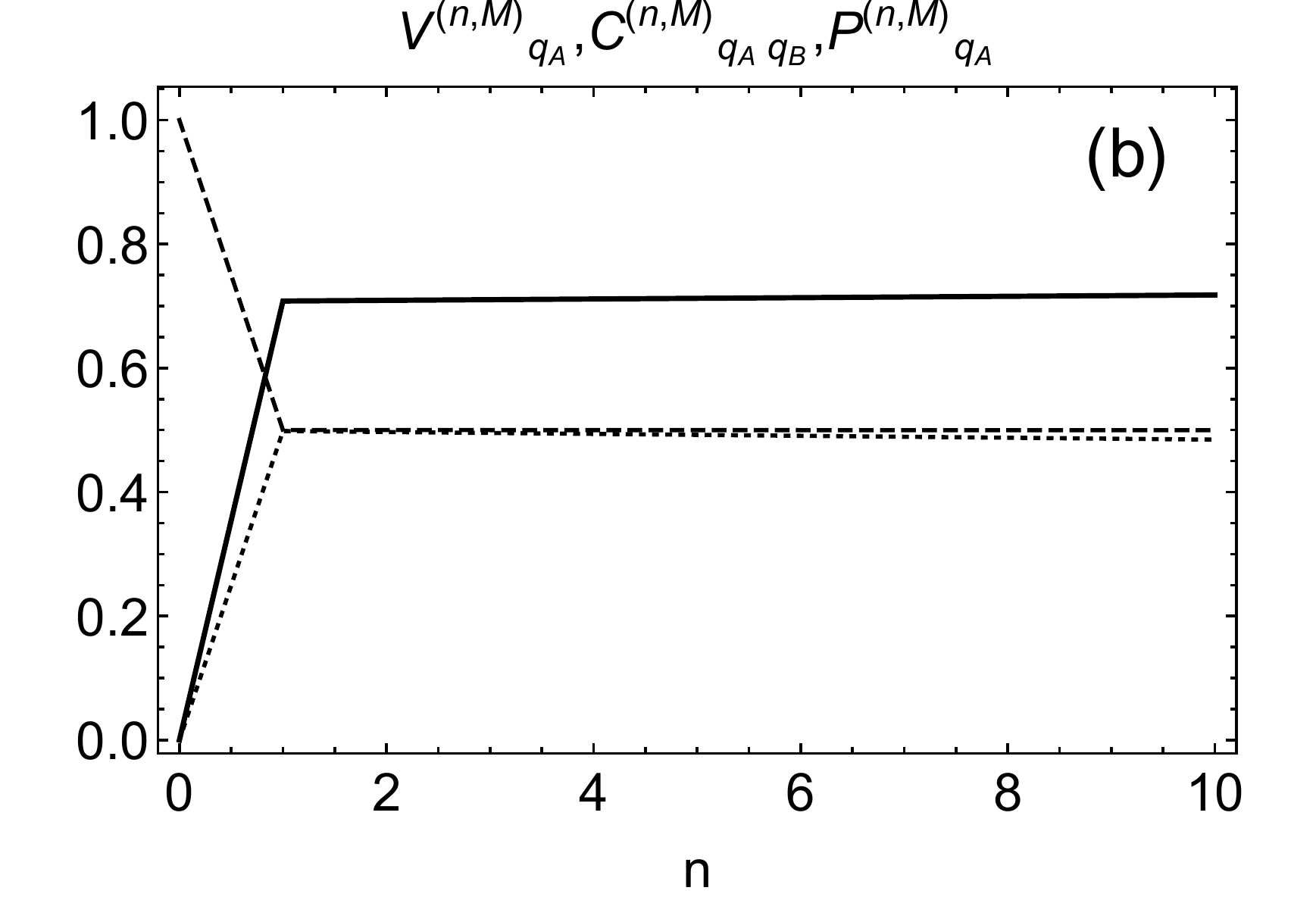} \label{maxvisb}}
\caption{(Color online) Complementarity quantities -- $V^{(n,M)}_{q_A}$ (solid), $C^{(n,M)}_{q_A,q_B}$ (dashed) and $P^{(n,M)}_{q_A}$ (dotted) -- as a function of \emph{n}, for optimization procedure in order to maximize the \emph{visibility}. Parameters: (a) $g T = 2 \pi \times 4$ and (b) $g T = \frac{2 \pi}{4}$. Also $N = 20$ and the coefficients $\alpha_i$ and $\beta_i$ are given by
the maximization procedure. In Figure (a), the solid red curve represents the limit $N \rightarrow \infty$, leading to $C_{q_A, q_B} = \ei^{\frac{-k T}{2}}$ (subsection \ref{digression} with $k=3$). All quantities are dimensionless.}
\label{maxvis}
\end{figure}

\begin{figure}[h]
\centering
{
\includegraphics[scale=0.42]{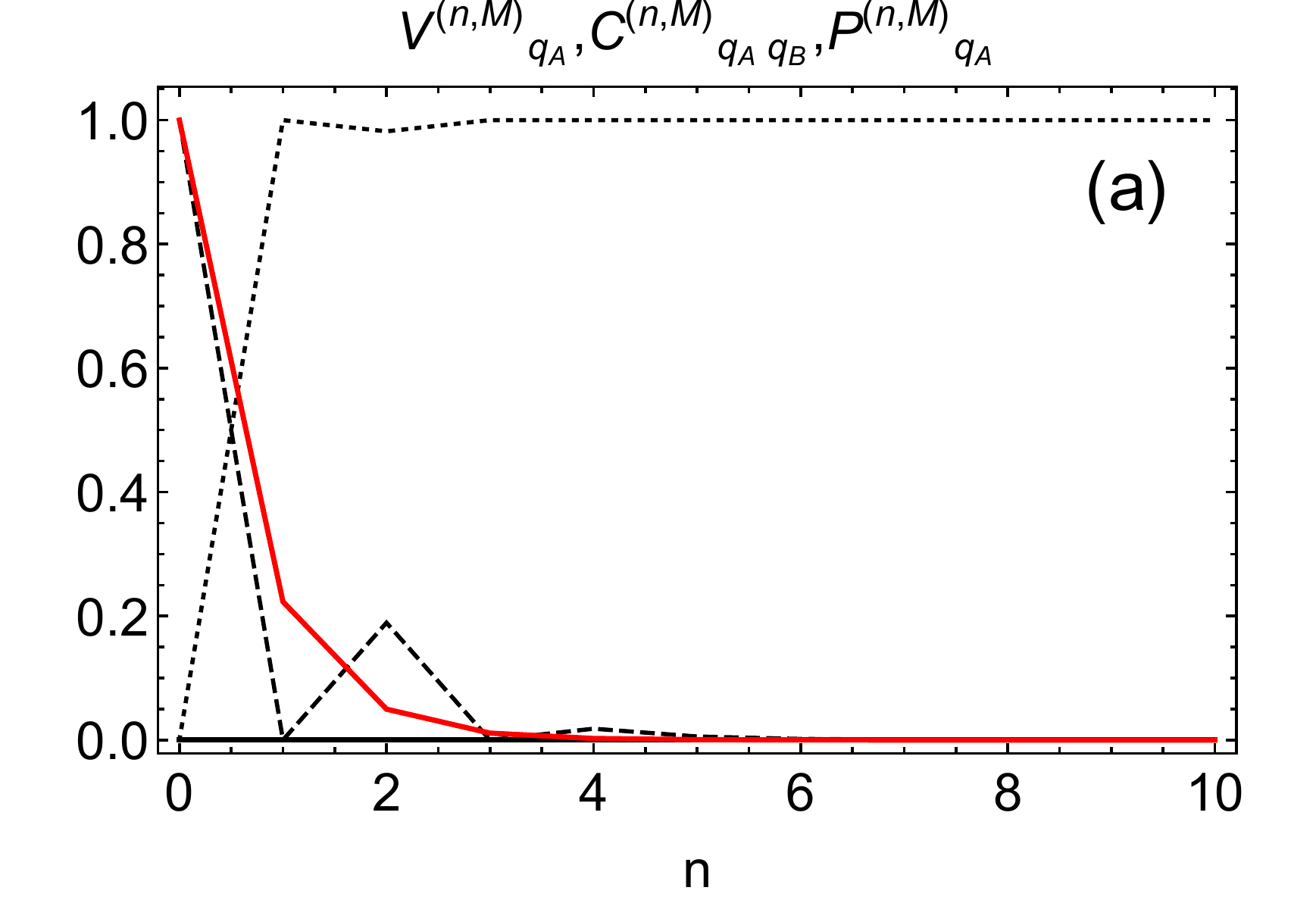} \label{maxprea}
}\hspace{0.5cm}
{
\includegraphics[scale=0.42]{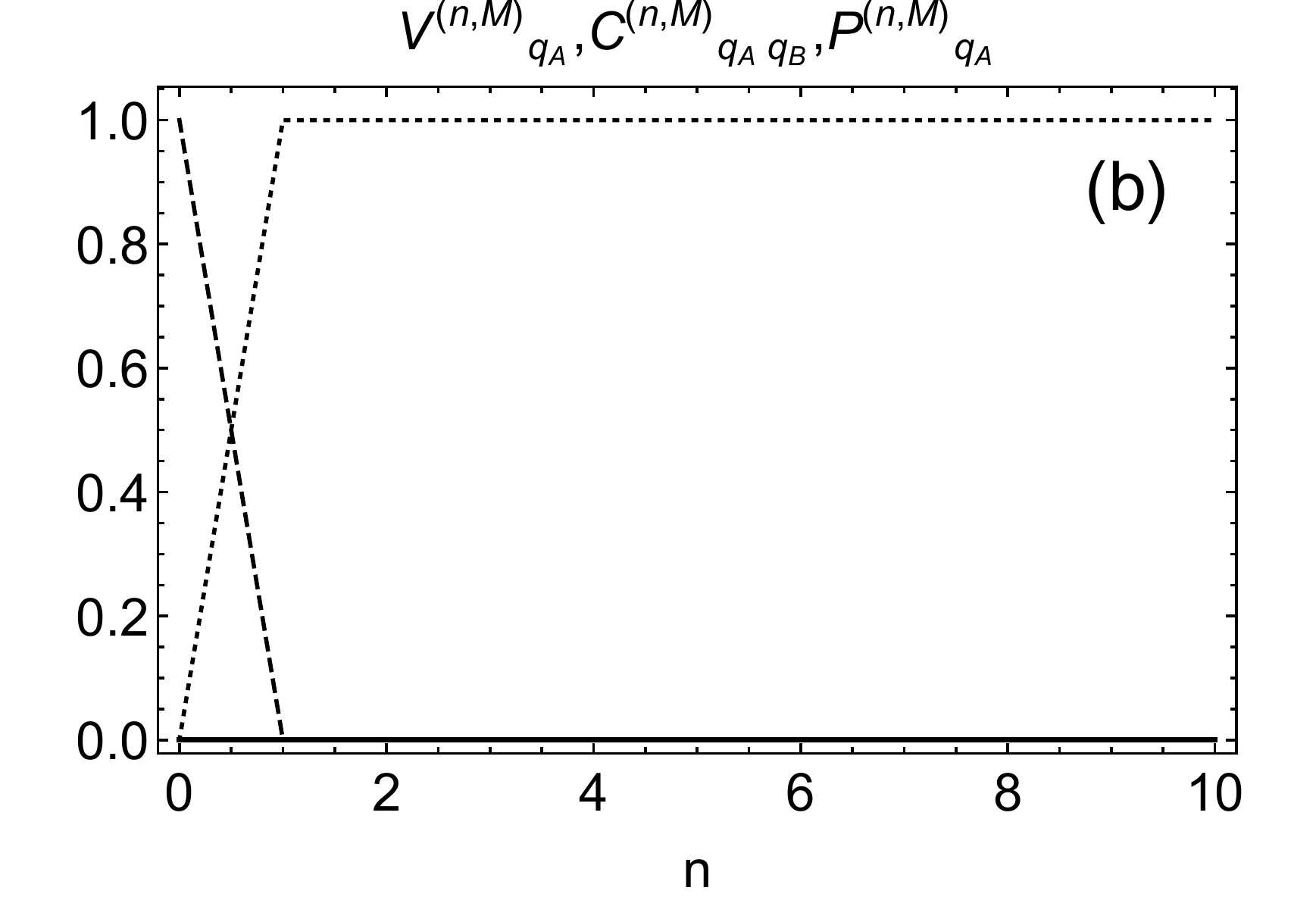} \label{maxpreb}
}
\caption{(Color online) Complementarity quantities -- $V^{(n,M)}_{q_A}$ (solid), $C^{(n,M)}_{q_A,q_B}$ (dashed) and $P^{(n,M)}_{q_A}$ (dotted) -- as a function of \emph{n}, for optimization procedure in order to maximize the \emph{predictability}. Parameters: (a) $g T = 2 \pi \times 4$ and (b) $g T = \frac{2 \pi}{4}$. Also $N = 20$ and the coefficients $\alpha_i$ and $\beta_i$ are given by
the maximization procedure. In Figure (a), the solid red curve represents the limit $N \rightarrow \infty$, leading to $C_{q_A, q_B} = \ei^{\frac{-k T}{2}}$ (subsection \ref{digression} with $k=3$). All quantities are dimensionless.}
\label{maxpre}
\end{figure}

\begin{figure}[h]
\centering
{\includegraphics[scale=0.42]{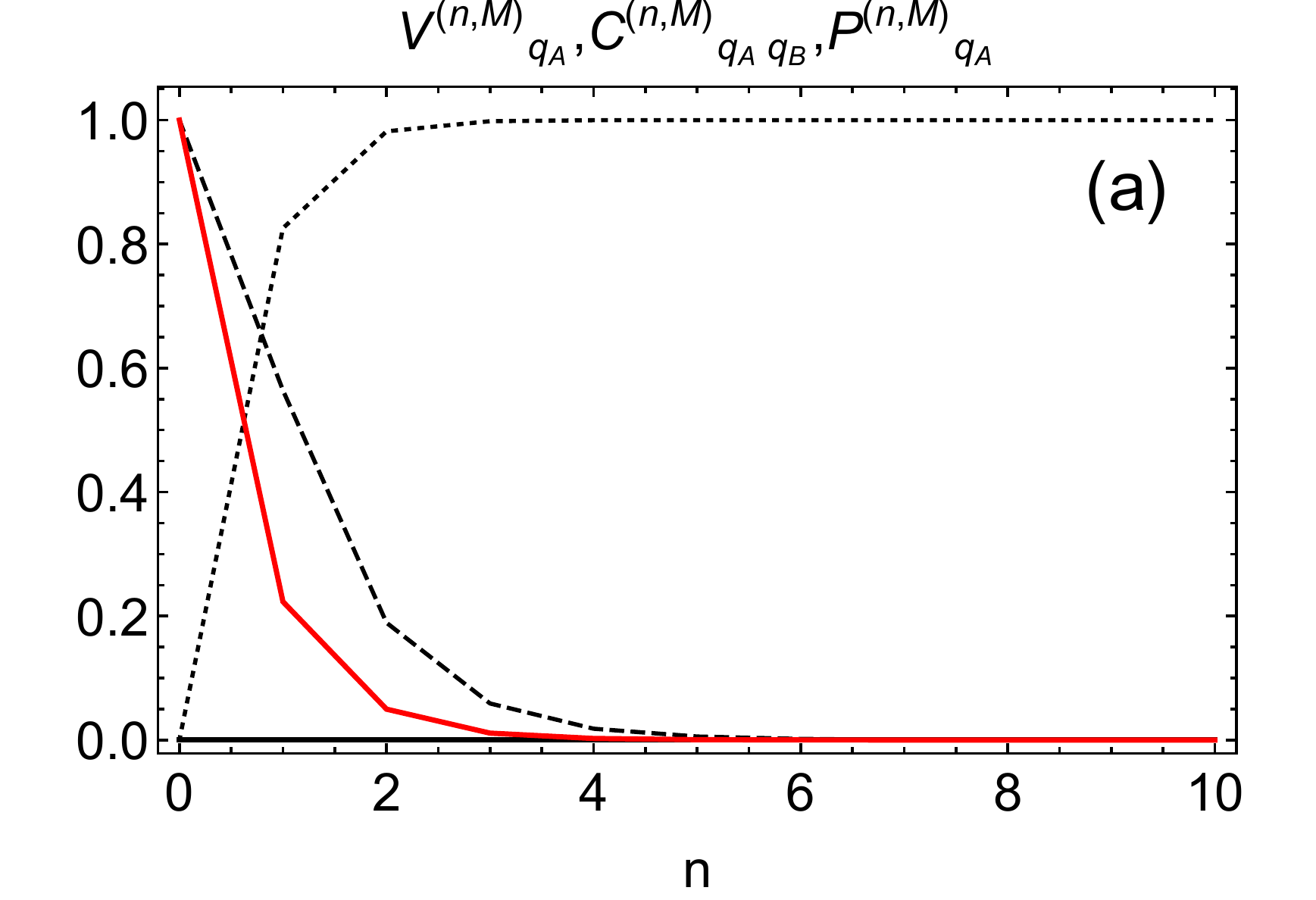}\label{maxcona}
}\hspace{0.5cm}
{
\includegraphics[scale=0.43]{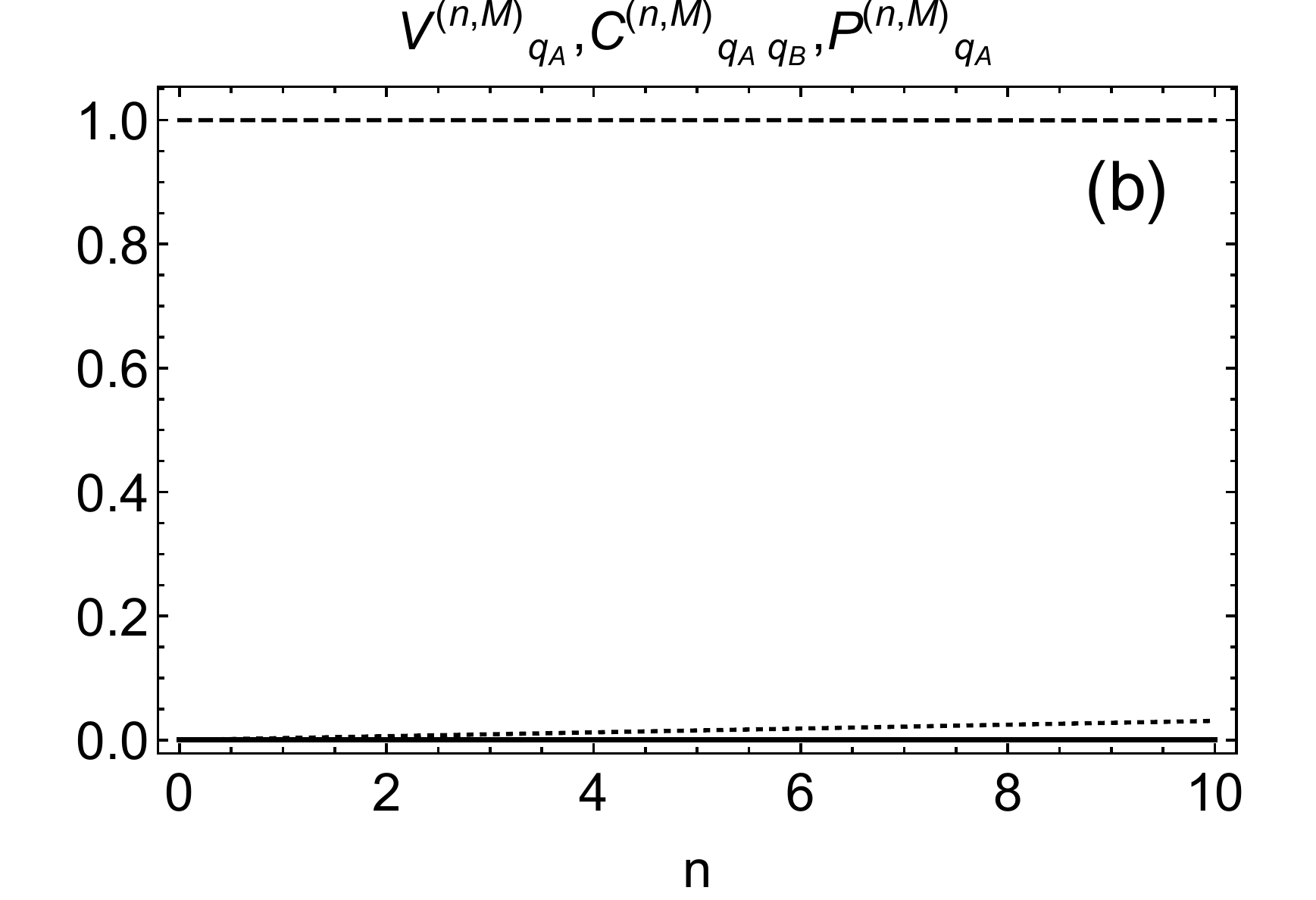}\label{maxconb}
}
\caption{(Color online) Complementarity quantities -- $V^{(n,M)}_{q_A}$ (solid), $C^{(n,M)}_{q_A,q_B}$ (dashed) and $P^{(n,M)}_{q_A}$ (dotted) -- as a function of \emph{n}, for optimization procedure in order to maximize the \emph{concurrence}. Parameters: (a) $g T = 2 \pi \times 4$ and (b) $g T = \frac{2 \pi}{4}$. Also $N = 20$ and the coefficients $\alpha_i$ and $\beta_i$ are given by
the maximization procedure. In Figure (a), the solid red curve represents the limit $N \rightarrow \infty$, leading to $C_{q_A, q_B} = \ei^{\frac{-k T}{2}}$ (subsection \ref{digression} with $k=3$). All quantities are dimensionless.}
\label{maxcon}
\end{figure}


In Figures \ref{maxvis}, \ref{maxpre} and \ref{maxcon} we show the Complementarity quantities, for the maximization of $V^{(n,M)}_{q_{A}}, P^{(n,M)}_{q_{A}}$ and $C^{(n,M)}_{q_{A}, q_B}$, respectively. We also present the behaviour for different couplings (each Figure (a) depicts $g = 4$ and each Figure (b) $g = \frac{1}{4}$). The \ankb{Black}{black} solid curve is related to the $\ankb{Visibility}{visibility}$; the same follows for the $concurrence$ (\ankb{Black}{black} dashed curve) and the $predictability$ (black dotted curve). Note in Figures \ref{maxvis}a and \ref{maxvis}b\ankb{, that for this coupling strength}{}, the \ankb{Visibility}{visibility} is \ankb{a}{an} increasing function of $n$, when the measurements are made to maximize $V^{(n,M)}_{q_{A}}$ \cite{Rossi2013}\ankb{, however}{. However,} if one \ankb{evaluates}{implements} measurements in order to maximize another Complementarity quantity, the \ankb{Visibility}{visibility} does not increase, and remains \ankb{in a value near}{close to} zero, as one can see in Figures \ref{maxpre} and \ref{maxcon}. \alams{Now, if the measurements are made in order to maximize the \emph{\ankb{Visibility}{visibility}}, for higher values of $g$, the behavior is completely different: the function goes to a maximum, and then decreases rapidly. In fact, since $(V^{(n,M)}_{q_{A}})^2+(C^{(n,M)}_{q_{A}, q_B})^2+(P^{(n,M)}_{q_{A}})^2 = 1$, if $V^{(n,M)}_{q_{A}}$ diminishes, other Complementarity quantity must increase.}{For smaller values of $g$, Figure \ref{maxvis}b, one can see that a perfect visibility is not reachable within our range of parameters, and this quantity reaches a maximum value $\sim 0.70$. This feature can be understood when we analyze the behavior of all the Complementarity quantities all together (Figures \ref{maxvis}b).}

\textcolor[rgb]{0.00,0.00,0.00}{Differently from the visibility, always when the predictability P is maximized it achieves the maximum value for some finite value of $n$ (Figures \ref{maxpre}a and \ref{maxpre}b). Moreover, it achieves a maximum valeu also when other quantities are maximized (Figure \ref{maxcon}a). Using an interferometric analogy, predictability is the which-way information that is available in the interferometric system, while \ankb{Visibility}{visibility} gives the quality of the interference pattern and concurrence is a measure of entanglement between $q_a$ (main system) and $q_B$ (which-way detector). Therefore, in Figures \ref{maxpre}a, \ref{maxpre}b and \ref{maxcon}a, since the system loses entanglement, with no acquisition of visibility, the predictability must increase.} \alams{}{} \alams{In Figures \ref{maxpre}a and \ref{maxpre}b it is possible to observe that the predictability tends to increase after some interactions, apart the case where the visibility is maximized (for $g T = 2 \pi \times 4$).}{} This feature can be understood in our approach, since the projective measurements will inherently modify the global system. In Table \ref{table} we show the states after $n$ interactions ($n = 1, 2$ and $10$), and after performing the maximization procedures. For instance, performing a maximization of $P_{q_A}^{(n,M)}$, the state $\ket{\psi}^{(n,M)}$ for $g T = 2 \pi \times 4$ tends to the state $\ket{1_A 0_B}$. \alams{}{For $g T = \frac{2 \pi}{4}$, the predictability can possess high values, as seen in Figure \ref{maxpre}b, since the asymptotic state is $\ket{0_A 0_B}$.}

\textcolor[rgb]{0.00,0.00,0.00}{The dashed curves in Figures \ref{maxvis}, \ref{maxpre} and \ref{maxcon} show the concurrence as a function of $n$. For $g T = \frac{2 \pi}{4}$, if the function to be maximized is the concurrence itself $C^{(n,M)}_{q_{A}, q_B}$ (Figure \ref{maxcon}b) \alams{or the predictability $P^{(n,M)}_{q_{A}}$}{}, it is possible to maintain the state almost maximally Entangled -- $C^{(n,M)}_{q_{A}, q_B} \sim 1$ -- by choosing the proper values of $\theta_i$ and $\phi_i$. \alams{}{Moreover, one can obtain Entanglement values near to $0.5$, performing measurements in order to maximize the visibility, Figure \ref{maxvis}b. This result is interesting, since we can see a clear complemental character between all quantities.} However, if the coupling is increased -- Figures \ref{maxvis}a, \ref{maxpre}a, or \ref{maxcon}a -- even if the projective measures were made to maximize the concurrence (Figure \ref{maxcon}a), Entanglement decreases to zero. This behavior is similar to two entangled qubits, where \alams{qubit}{one of them, say} $q_B$, is coupled to a thermal reservoir (red solid curves), but in our case we have a finite number of interacting qubits with $q_B$ (where the maximum number of interacting qubits is $N = 20$). \alams{This feature can be understood when we analyse}{This can be conprehended by analyzing} how the initial information (given by the distinguishability between $q_A$ and the $i$-th qubit) is distributed over the $N$ qubits \alams{}{(section \ref{distsec})}. An interesting aspect concerning Figure \ref{maxcon}b, for $g T = \frac{2 \pi}{4}$, is that one can see an approximately steady behavior of the concurrence $C^{(n,M)}_{q_{A}, q_B}$, near the initial value $C^{(0)}_{q_{A}, q_B} = 1$. It is possible, therefore, to maintain the system $q_A + q_B$ in an approximately maximal Entangled state, notwithstanding the qubits of $R$ became dynamically correlated with $q_A$ and $q_B$ (Equation \eqref{rhon}). One can see from Table \ref{table}, for $g T = \frac{2 \pi}{4}$ and $n=10$, the state resemble the initial maximally entangled state\alams{, for all maximization procedures,}{, if the maximization is over $C_{q_A, q_B}^{(n,M)}, $} corroborating Figure \ref{maxcon}b.}

\begin{table*}[t]
\caption{Approximate values of the state $\ket{\psi}^{(n,M)}$ after $n$ measurements, for different optimization procedures and coupling strengths.}\label{table}
\begin{center}
\begin{ruledtabular}
\begin{tabular}{cc||c||cc}
& & $\mathbf{gT = 2 \pi \times 4}$& $\mathbf{gT = 2 \pi / 4}$ &   \\[5pt] \cline{1-4}
\multicolumn{1}{ c  }{\multirow{3}{*}{\textbf{Max. of $P_{q_A}^{(n,M)}$}} } &
\multicolumn{1}{ c|| }{$\mathbf{n=1}$} & $\Big( (-0.05 + 0.99 i) \ket{0_A 0_B} \Big)\otimes \ket{M}$ & $\Big( (-0.73 + 0.68 i) \ket{0_A 0_B} \Big)\otimes \ket{M}$ &     \\[5pt] 
\multicolumn{1}{ c  }{}                        &
\multicolumn{1}{ c|| }{$\mathbf{n=2}$} & $\Big( 0.10 \ket{0_A 1_B} + 0.99 \ket{1_A 0_B} \Big)\otimes \ket{M}$ & $\Big( (-0.45 + 0.89 i) \ket{0_A 0_B} \Big)\otimes \ket{M}$ &       \\[5pt] 
\multicolumn{1}{ c  }{}                        &
\multicolumn{1}{ c|| }{$\mathbf{n=10}$} & $\Big(  \ket{1_A 0_B} \Big)\otimes \ket{M}$ & $\Big( (-0.15 + 0.98 i) \ket{0_A 0_B} \Big)\otimes \ket{M}$ &       \\[5pt] \cline{1-4}
\multicolumn{1}{ c  }{\multirow{3}{*}{\textbf{Max. of $V_{q_A}^{(n,M)}$}} } &
\multicolumn{1}{ c|| }{$\mathbf{n=1}$} & $\Big( (0.23 - 0.66 i) \ket{0_A 0_B} + 0.21 \ket{0_A 1_B} + 0.67 \ket{1_A 0_B} \Big)\otimes \ket{M}$ & $\Big( (0.61 - 0.35 i) \ket{0_A 0_B} - 0.50 \ket{0_A 1_B} - 0.50 \ket{1_A 0_B} \Big)\otimes \ket{M}$ &     \\[5pt] 
\multicolumn{1}{ c  }{}                        &
\multicolumn{1}{ c|| }{$\mathbf{n=2}$} & $\Big( (0.17 - 0.68 i) \ket{0_A 0_B} + 0.06 \ket{0_A 1_B} + 0.70 \ket{1_A 0_B} \Big)\otimes \ket{M}$ & $\Big( (0.47 - 0.52 i) \ket{0_A 0_B} + 0.49 \ket{0_A 1_B} + 0.49 \ket{1_A 0_B} \Big)\otimes \ket{M}$ &       \\[5pt] 
\multicolumn{1}{ c  }{}                        &
\multicolumn{1}{ c|| }{$\mathbf{n=10}$} & $\Big( (-0.43 + 0.55 i) \ket{0_A 0_B} + 0.70 \ket{1_A 0_B} \Big)\otimes \ket{M}$ & $\Big( (0.40 - 0.58 i) \ket{0_A 0_B} - 0.49 \ket{0_A 1_B} - 0.50 \ket{1_A 0_B} \Big)\otimes \ket{M}$ &       \\[5pt] \cline{1-4}
\multicolumn{1}{ c  }{\multirow{3}{*}{\textbf{Max. of $C_{q_A,q_B}^{(n,M)}$}} } &
\multicolumn{1}{ c|| }{$\mathbf{n=1}$} & $\Big( 0.29 \ket{0_A 1_B} + 0.95 \ket{1_A 0_B} \Big)\otimes \ket{M}$ & $\Big( 0.71 \ket{0_A 1_B} + 0.71 \ket{1_A 0_B} \Big)\otimes \ket{M}$ &     \\[5pt] 
\multicolumn{1}{ c  }{}                        &
\multicolumn{1}{ c|| }{$\mathbf{n=2}$} & $\Big( 0.10 \ket{0_A 1_B} + 0.99 \ket{1_A 0_B} \Big)\otimes \ket{M}$ & $\Big( 0.70 \ket{0_A 1_B} + 0.71 \ket{1_A 0_B} \Big)\otimes \ket{M}$ &       \\[5pt] 
\multicolumn{1}{ c  }{}                        &
\multicolumn{1}{ c|| }{$\mathbf{n=10}$} & $ \Big(\ket{1_A 0_B} \Big)\otimes \ket{M}$ & $ \Big( 0.69 \ket{0_A 1_B} + 0.72 \ket{1_A 0_B} \Big)\otimes \ket{M}$ &       \\[5pt] 
\end{tabular}
\end{ruledtabular}
\end{center}
\end{table*}

\subsection{Complementarity without maximization procedure}\label{noMaximization}

\begin{figure*}[t]
\centering
\includegraphics[scale=0.3]{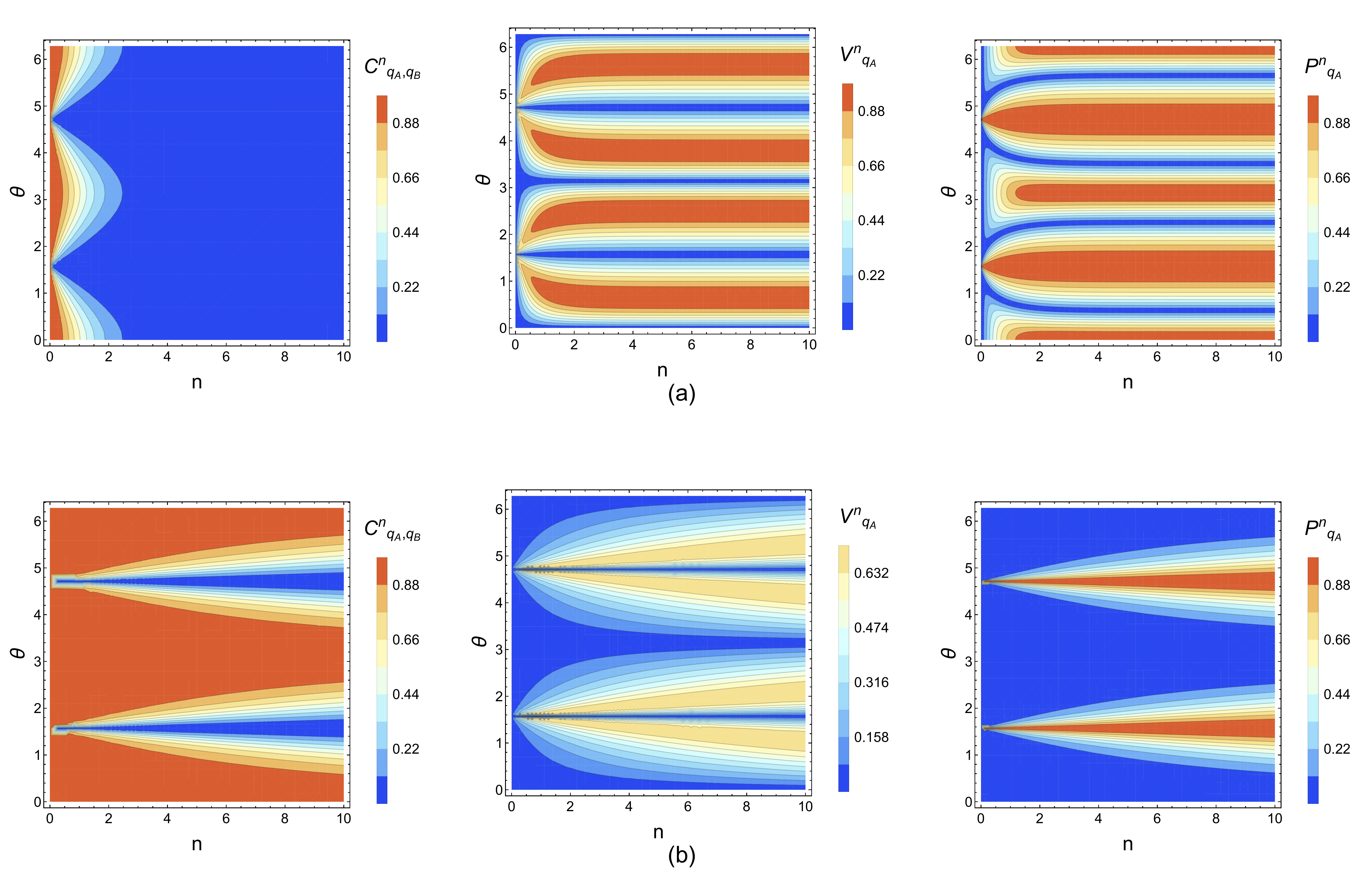}
\caption{(Color online) $C^{(n,M)}_{q_{A}, q_B}$, $V^{(n,M)}_{q_{A}}$ and $P^{(n,M)}_{q_{A}}$, respectively, as a function of $n$ and $\theta$. Parameters: $N = 20$, (a) $g T = 2 \pi \times 4$ and (b) $g T = \frac{2 \pi}{4}$. All quantities are dimensionless.}\label{alfaequal}
\end{figure*}

\alams{}{Here we develop a study without assuming any maximization, adding an evident picture of the complementarity between the quantities. }Suppose that instead of performing the measurements $\Pi_i$ in order to maximize a given quantity, one is able to project the state $\rho^{(n)}$ only in the same base for each qubit of $R$. In other words, lets consider the $i$-qubit of $R$ is projected in the state $\ket{M_i} = \alpha \ket{0_i} + \beta \ket{1_i}$, where $\alpha = \cos \theta$ and $\beta = \ei^{i \phi} \sin \theta$ are \alams{equal}{the same, individually, } for all measurements. With this assumption, the coefficients for the state \eqref{rhoreducedM} are given by: \bq \gamma_1 &=& \frac{(a^n -1) b \alpha^n \beta}{\sqrt{2} (a-1) \alpha}, \nonumber \\ \gamma_2 &=& \frac{\alpha^n a^n}{\sqrt{2}}, \nonumber \\ \gamma_3 &=& \frac{\alpha^n}{\sqrt{2}}, \eq where $a = \cos \left( \frac{g T}{N}\right)$ and $b = - i \sin \left( \frac{g T}{N}\right)$. The coefficients above allow us to calculate the Complementarity quantities $V^{(n,M)}_{q_{A}}, P^{(n,M)}_{q_{A}}$ and $C^{(n,M)}_{q_{A}, q_B}$, that we depict in Figure \ref{alfaequal} for different couplings strength. One can see clearly the Complementarity behavior between the quantities in all plots.

Note in Figure \ref{alfaequal}a ($g T = 2 \pi \times 4$), when $\theta = \frac{m \pi}{2}$ ($m \in$ Integer) for large values of $n$, the predictability has values near unity, while the \ankb{Visibility}{visibility} is near zero (the concurrence is negligible in this regime, as pointed out previously). This can be seen through Equation \eqref{rhon}: if $\theta = \frac{m \pi}{2}$ for odd values of $m$ the state after $n$ measurements is approximately $\rho_{m} \thickapprox \ket{1_A 0_B}$, while for $m$ assuming even values the state will become $\rho_{m} \thickapprox \ket{0_A 0_B}$, explicitly the maximum values of the predictability in Figure \ref{alfaequal}a. The same argument can be used for the \ankb{Visibility}{visibility} in this regime, but in this case $\theta = \frac{(2m+1) \pi}{4}$ ($m$ Integer) and the final state tends to $\rho_{m} \thickapprox \frac{\ket{1_A 0_B} + \ket{0_A 1_B}}{\sqrt{2}}$.

Figure \ref{alfaequal}b, on the other hand, shows the Complementarity quantities in the weak coupling regime. It is noticeable that one can sustain the state in the maximal entangled state by performing specific measurements ($\theta = m \pi$, with $m$ Integer). This can be achieved since for this coupling regime only little information about the initial state is captured by the interaction between $q_B$ and $q_i$ (Subsection \ref{distsec}). The large values for the predictability (and consequently small values of \ankb{Visibility}{visibility} and concurrence), when $\theta = \frac{m \pi}{2}$ (where $m$ is an odd Integer), can be understood in the same sense as the case of $g T = 2 \pi \times 4$ (Figure \ref{alfaequal}a), where the state is approximately $\rho_{m} \thickapprox \ket{1_A 0_B}$.

\subsection{Information distribution - Distinguishability}\label{distsec}

In this section we use the distinguishability to quantify the information stored by each subsystem. We show how this quantity varies as one evaluate the measurements. We are concerned in how the information stored in some parts of the global system behaves, given that $n$ measurements are performed in the subsystem $R$. The distinguishability is calculated before and after $n$ measurements, and we show the curves that illustrate the behavior of this quantity for the maximizations of $V$, $P$ and $C$.

\begin{figure}[h]
\centering
{
\includegraphics[scale=0.4]{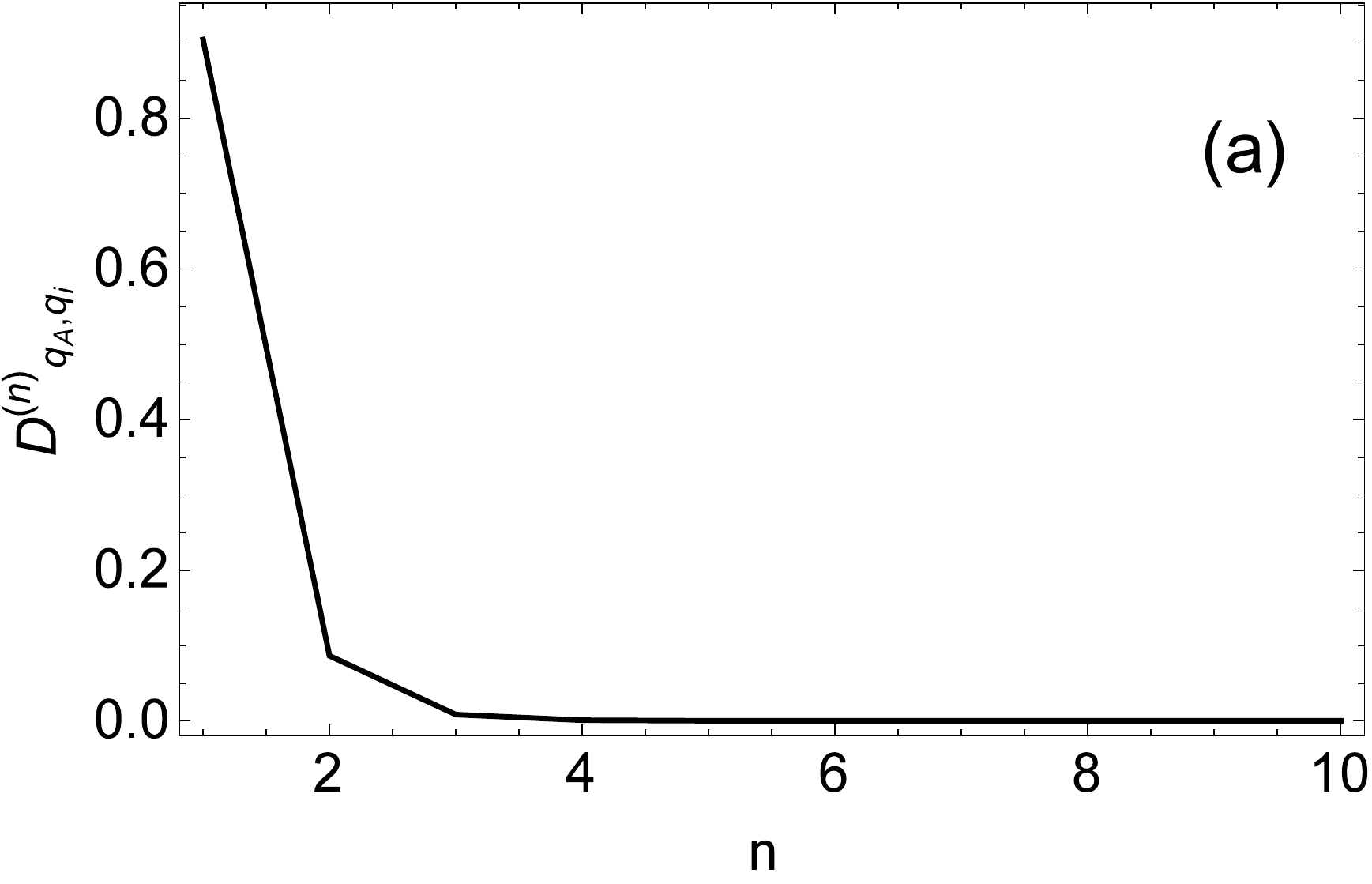}\label{dist_a}
}
{
\includegraphics[scale=0.4]{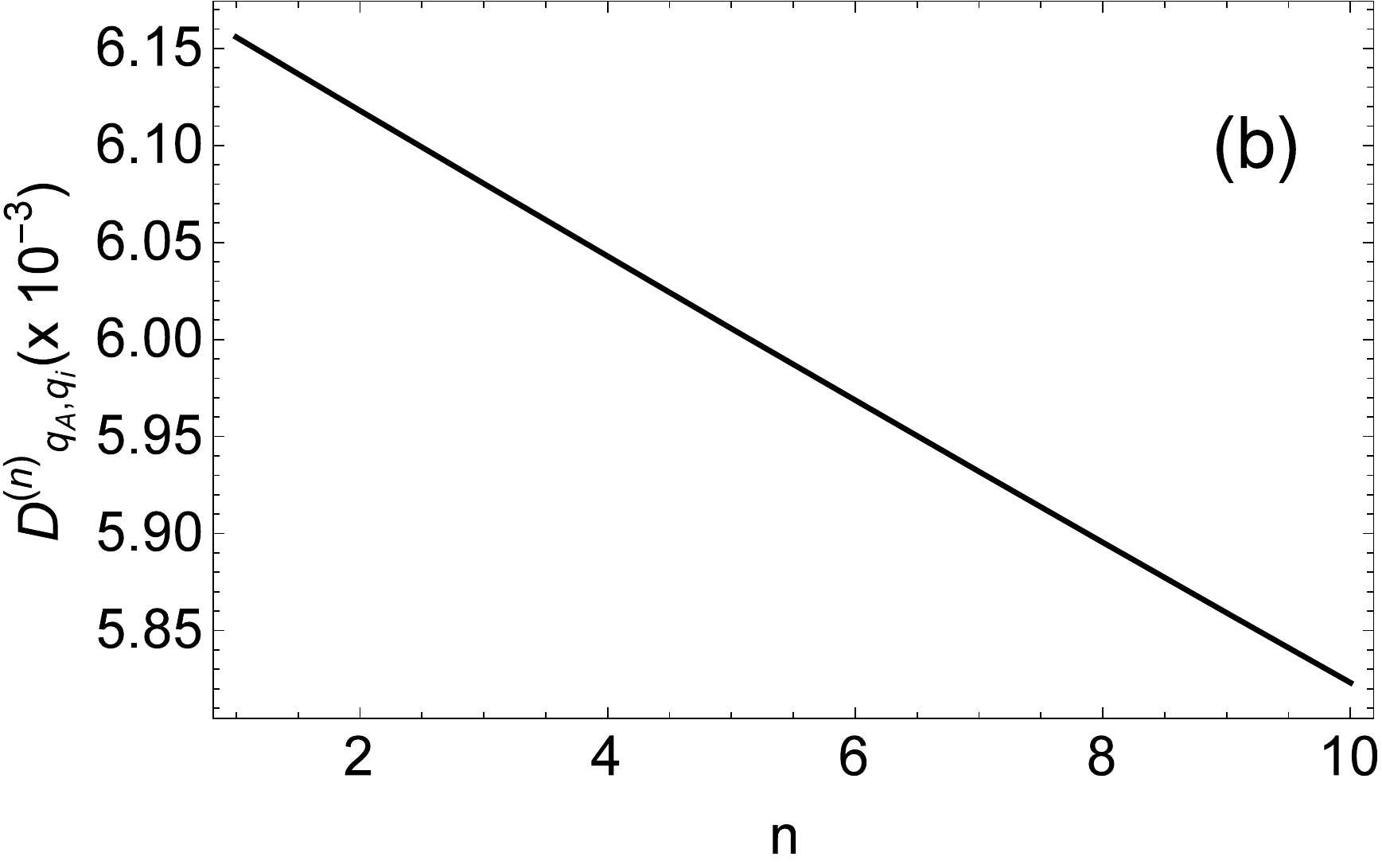}\label{dist_b}
}
\caption{Distinguishability between $q_A$ and the $i$-th qubit of $R$, as a function of \emph{n}, i.e. how much information about the initial state the qubit $i$ have. Parameters: $N = 20$, (a) $g T = 2 \pi \times 4$ and (b) $g T = \frac{2 \pi}{4}$. All quantities are dimensionless.}
\label{dist}
\end{figure}

Figures \ref{dist}a and \ref{dist}b show how the distinguishability is distributed in the $N$ qubits, i.e. $D^{(n)}_{q_A,q_i}$: \bq D_{q_{A},q_i}^{(n)} &=& \tr_{q_i} \Big\{ \Big|
\frac{1}{2} a^{2n} \ket{0_i} \bra{0_i} + \abs{a^{i-1} b}^2 \ket{1_i}
\bra{1_i} + \nonumber \\ && + (\abs{a^{n-1} b}^2 + \ldots  + \abs{a^{i-2} b}^2 + \nonumber \\ &&+
\abs{a^i b} + \ldots + \abs{b}^2) \ket{0_i}
\bra{0_i} - \ket{0_i} \bra{0_i} \Big| \Big\}  = \nonumber \\ &=& \abs{a^{i-1} b}^2. \nonumber \\ \label{dist_eq}\eq This relation gives the amount of information the $i$-th qubit possess about the initial state. Both Figures are independent of the maximization procedure. Note that for $g T = \frac{2 \pi}{4}$, Figure \ref{dist}b, the information is almost equally distributed in all the qubits of $R$. For coupling $g T = 2 \pi \times 4$, the first qubits that interacted with $q_B$ retain a large amount of information from $q_A + q_B$ (Figure \ref{dist}a). Comparing the results of Figures \ref{maxcon}a and \ref{dist}a, we can argue that, if $g T = 2 \pi \times 4$ (for high coupling intensity between $R$ and $q_B$), the first qubits of $R$ that interact with $q_B$ ``extract'' sufficient information, that was initially stored only in $q_A + q_B$. Therefore, the global system ($R + q_A + q_B$) becomes strongly correlated, leading the concurrence between $q_A$ and $q_B$ diminishes rapidly, compared with the case $g T = \frac{2 \pi}{4}$ (Figures \ref{maxvis}a, \ref{maxpre}a and \ref{maxcon}a).

\begin{figure}[h]
\centering
{
\includegraphics[scale=0.4]{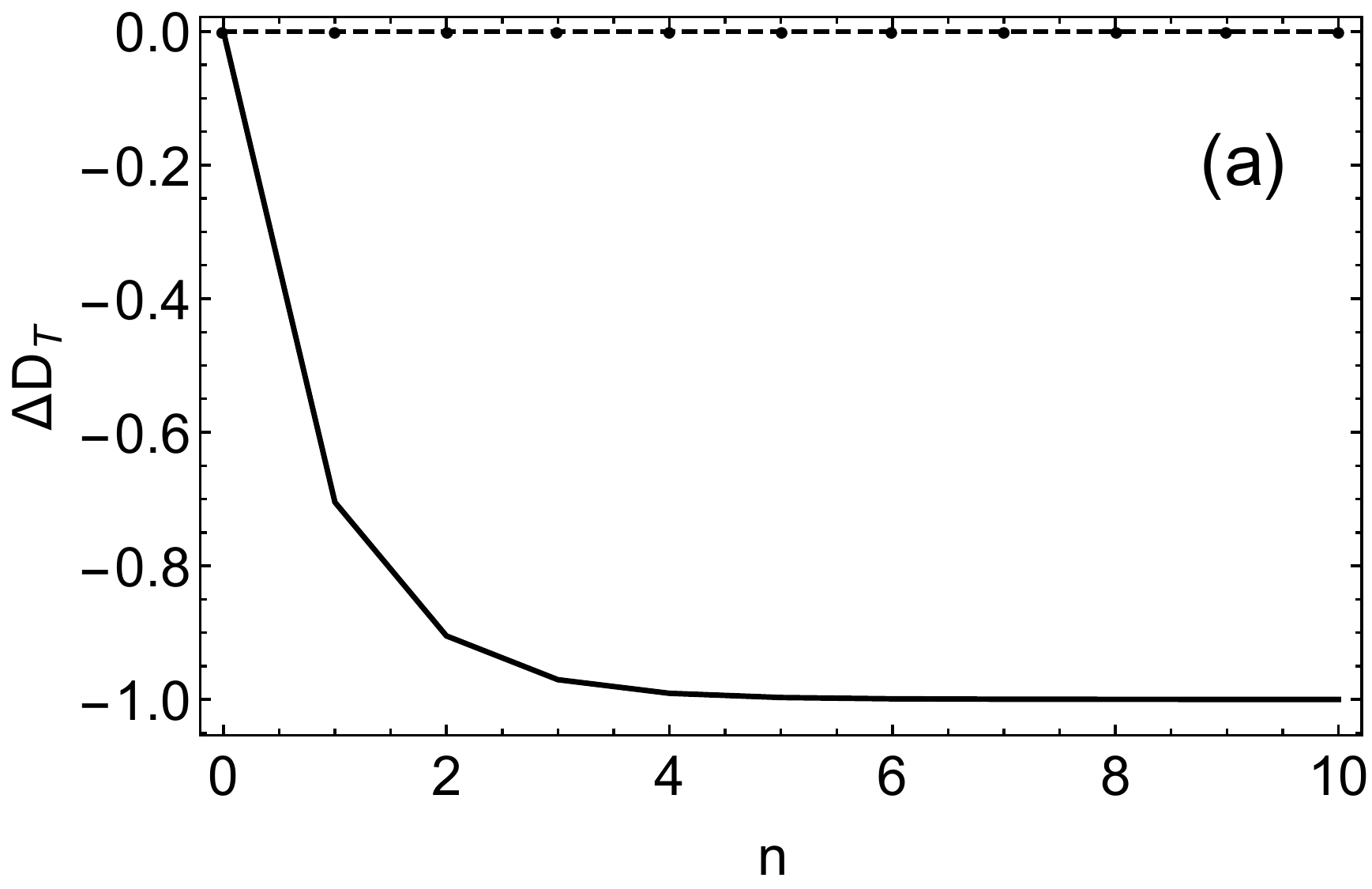}\label{maxdeltaTa}
}
{
\includegraphics[scale=0.4]{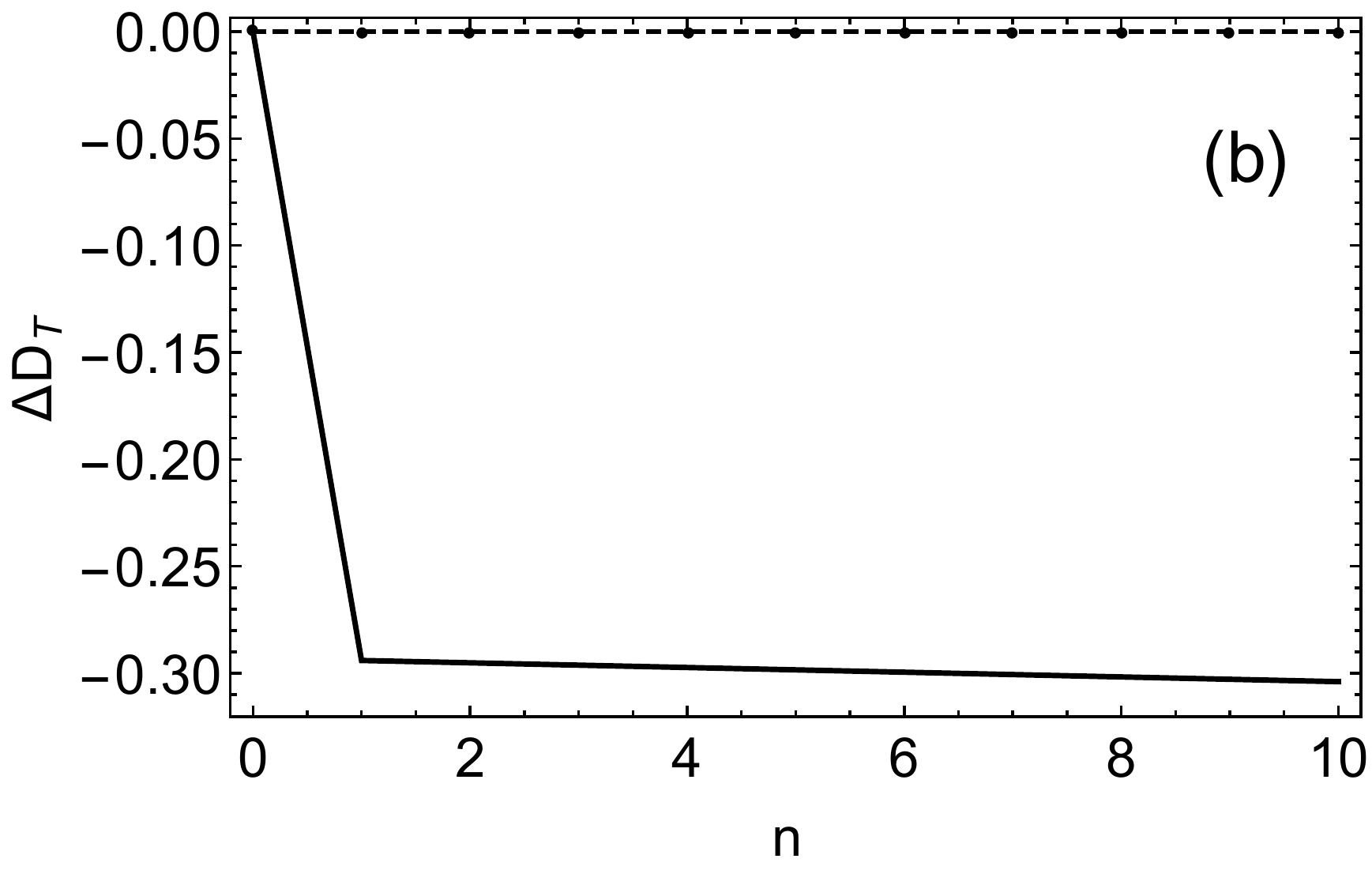}\label{maxdeltaTb}
}
\caption{$\Delta D_T$ as a function of \emph{n}, for optimization procedure in order to maximize the \ankb{Visibility}{visibility} (solid black), the concurrence (dashed black) and the predictability (dotted black). Parameters: (a) $g T = 2 \pi \times 4$ and (b) $g T = \frac{2 \pi}{4}$. Also, $N = 20$ and the coefficients $\alpha_i$ and $\beta_i$ are given by
the maximization procedure. All quantities are dimensionless.}
\label{maxdeltaT}
\end{figure}


The total Distinguishability, for the global system composed by $q_A + q_B + R$, is:
\bq D^{(n)}_{T} &=& \sum^{N}_{i\neq q_{A}}C^{2}_{q_{A},i} = a^{2n}+ \abs{a^{n-1} b}^2 + \ldots +
\abs{a^i b}^{2} + \ldots + \abs{b}^2 \nonumber \\ &=& D_{q_{B}}^{(n)} +
\sum_{i=1}^{n} D_{q_i}^{(n)} = 1. \eq After performing the projective measurements, the Distinguishability of the subsystem $q_{A} + q_{B}$ is:
\be D^{(n,M)}_{T}= \sqrt{\left(C^{(n,M)}_{q_{A},q_{B}}\right)^{2}+\left(P^{(n,M)}_{q_{A}}\right)^{2}}.
\ee Since the projected subsystem can now be decoupled from $q_A$ and $q_B$ \eqref{rhoreducedM}, the variation of the total Distinguishability is: \bq \Delta D_{T} &=& D^{(n,\textbf{M})}_{T} - D^{(n)}_{T} =\sqrt{\left(C^{(n,\textbf{M})}_{q_{A},q_{B}}\right)^{2}+\left(P^{(n,\textbf{M})}_{q_{A}}\right)^{2}} - 1 \nonumber \\ &=& \sqrt{1-\left(V^{(n,\textbf{M})}_{q_{A}}\right)^{2}} - 1.
\eq The variation of Distinguishability between subsystems $q_{A}$ and $q_{B}$ is given by \eqref{DnqAqB}: \be \Delta D_{F} = D^{(n,\textbf{M})}_{T} - D^{(n)}_{q_A, q_B} = \sqrt{1-\left(V^{(n, M)}_{q_{A}}\right)^{2}}-a^{2n}.\ee

Figures \ref{maxdeltaT}a and \ref{maxdeltaT}b show $\Delta D_T$ for \ankb{$g T = 2 \pi \times 4$}{$g T = 2 \pi \times 4$} and \ankb{$g T = \frac{2 \pi}{4}$}{$g T = \frac{2 \pi}{4}$}, respectively. We can see that the measurements that maximizes the visibility, erases the information stored by the qubits of $R$. Notice that in \ref{maxdeltaT}a for $N=4$ the the information of the global system is approximately zero, this is in accord with Figure \ref{maxvis}a that shows visibility approximately $1$ for $N=4$.  In figure \ref{maxdeltaT}b the stabilization of the curve shows that the information erased is limited, because it is equally distributed among the qubits of $R$, as it is shown in Figure \ref{dist}b, therefore it is erased by small amounts after each measurements. Measurements that maximizes predictability and concurrence show no variation of information before and after measurements of the qubits, therefore, they do not erase information form the system $R$.


%

\section{Conclusion}\label{conclusion}

In this work we have proposed and discussed in details a scheme to observe the behavior of Complementarity quantities (concurrence, \ankb{Visibility}{visibility} and predictability) of a two qubit system, initially maximally entangled, after two steps: one of the qubits ($q_B$) interacts with $R$ (composed by $N$ other qubits -- $q_i$); after the interaction, projective measurements are made in each qubit of $R$ in order to maximize a given quantity. We observe that, if the coupling strength between $q_B$ and $R$ is considerable, the concurrence behaves similarly as a system of two qubits coupled to a thermal reservoir, even though our subsystem $R$ is composed by a finite number of qubits $N$. On the other hand, if the coupling is ``small'' (compared to the previous one), the system entanglement may as well be preserved. \textcolor[rgb]{0.00,0.00,0.00}{The visibility however shows a different behavior, when the coupling is stronger its maximization is more effective.} To explicit these results, we show some intermediate states for different couplings and number of interactions. The differences of the behavior can be understood from the distribution of information over the global system, as measured by the Distinguishability between $q_A+q_i$. We also studied the variation of distinguishability (before and after the measurements) for the global system and for the initial qubits $q_A + q_B$, making a connection between the information stored in each part of the system and the corresponding behavior of the Complementarity quantities. Note that the presented model may be feasible experimentally. One can, in principle, call qubits $q_A$ and $q_B$ as the modes inside microwave cavities, which nowadays are constructed with a lifetime of approximately $0.1$ seconds \cite{Gleyzes2007, Kuhr2007}. The qubits of the subsystem $R$ could represent two-level atoms in a QED experiment, for instance. The interaction time between each atom ($q_i$) and the mode $q_B$ is of the order of $10^{-5} s$ \cite{Brune1996}. So, for the $10$ qubits of our model, the effect would be in fact visible and dissipation can be neglected.

We showed how to fully control and prepare maximized states for different complementarity quantities. Since many parts are involved in this control scheme, this could have applications for protocols of quantum information, for example as protocols of bit commitment \cite{bitcomm1, bitcomm2}, where Alice wants to save safely the information of her bit for some time, but wants to reveal it later on to Bob. In this case, we could imagine that Alice would have access to part R and Bob, on the other hand, would have access to qubit A  and B.

\begin{acknowledgments}
The authors would like to thank the support from the Brazilian agencies CNPq and CAPES (CAPES (6842/2014-03); CNPq (470131/2013-6)). L.A.M.S. also thanks the University of Nottingham for hospitality and support during part of this work preparation. The authors acknowledge useful discussions with P. Saldanha, M. P. Fran\c{c}a Santos and G. Murta.
\end{acknowledgments}

\end{document}